\documentclass[apj]{emulateapj}
\usepackage{graphicx,subfigure,enumerate,pstricks,latexsym,longtable}
\usepackage[english]{babel}

\newcommand{\Schw}{Schwarzschild}

\newcommand{\beq}{\begin{equation}}
\newcommand{\eeq}{\end{equation}}
\newcommand{\bea}{\begin{eqnarray}}
\newcommand{\eea}{\end{eqnarray}}

\newcommand{\ce}{{\cal{E}}} 
\newcommand{\cl}{{\cal{L}}} 
\newcommand{\cb}{{\cal{B}}}
\newcommand{\BB}{{\cal{B}}}
\newcommand{\KK}{{\cal{K}}}

\newcommand{\q}{ {\tilde{q}} }

  \def\cp{\pi}

\providecommand{\dif}{\mathrm{d}} \def\d{\dif}

\def\mir{\mathrm{r}}
\def\mit{\mathrm{\theta}}
\def\mip{\mathrm{\phi}}
\def\mil{\mathrm{L}}

\def\aprx{\sim}

\def\pre{Phys. Rev. E }
\def\prd{Phys. Rev. D }
\def\jcap{Journal of Cosmology and Astroparticle Physics }

\def\mnras{Monthly Notices of the Royal Astronomical Society }
\def\apj{The Astrophysical Journal }
\def\aap{Astronomy and Astrophysics }

\def\nat{Nature }

\begin{document}

\title{Radiation reaction of charged particles orbiting \\
magnetized Schwarzschild black hole}

\author{Arman Tursunov$^{(1)}$}
\author{Martin Kolo\v{s}$^{(1)}$}
\author{Zden\v{e}k Stuchl{\'i}k$^{(1)}$}
\author{Dmitri V. Gal'tsov$^{(2,3)}$}

\affiliation{$^{(1)}$Institute of Physics and Research Centre of Theoretical Physics and Astrophysics,\\
 Silesian University in Opava, Bezru{\v c}ovo n{\'a}m.13, CZ-74601 Opava, Czech Republic}
\affiliation{$^{(2)}$Faculty of Physics, Moscow State University, 
119899, Moscow, Russia}
\affiliation{$^{(3)}$Kazan Federal University, 420008 Kazan, Russia }

\begin{abstract}
In many astrophysically relevant situations radiation reaction force acting upon a charge can not be neglected and the question arises about the location and stability of circular orbits in such regime. Motion of point charge with radiation reaction in flat spacetime is described by Lorenz-Dirac (LD) equation, while in curved spacetime -- by DeWitt-Brehme (DWB) equation containing the Ricci term and the tail term. We show that for the motion of elementary particles in vacuum metrics the DWB equation can be reduced to the covariant form of the LD equation which we use here. Generically, the LD equation is plagued by runaway solutions, so we discuss computational ways to avoid this problem in constructing numerical solutions. We also use the first iteration of the covariant LD equation which is the covariant Landau-Lifshitz equation, comparing results of these two approaches and showing smallness of the third-order Schott term in the ultrarelativistic case. We calculate the corresponding energy and angular momentum loss of a particle and study the damping of charged particle oscillations around an equilibrium radius. We find that depending on the orientation of the Lorentz force, the oscillating charged particle either spirals down to the black hole, or stabilizes the circular orbit by decaying its oscillations. The later case leads to an interesting new result of shifting of the particle orbit outwards from the black hole. We also discuss the astrophysical relevance of the presented approach and provide estimations of the main parameters of the model.
\end{abstract}


 

\maketitle


\section{Introduction} \label{sec-intro}

Synchrotron radiation emitted by a charged particle leads to appearance of the back-reaction force which can significantly affect its motion. The purpose of this paper is to study the motion of charged particles in the combined magnetic and gravitational fields, taking into account the radiation reaction force.
There is convincing evidence that magnetic fields are indeed present in the vicinity of black holes. The observations of the Galactic Centre (\cite{Eatough-etal:2013:Natur:}) have demonstrated the existence of strong magnetic field of hundred Gauss in the vicinity of the supermassive black hole at the Galactic centre. Recent studies of the quasi-periodic oscillations (QPOs) observed in the black hole microquasars have shown possible signatures of Galactic magnetic fields in the vicinity of three microquasars (\cite{Kol-Tur-Stu:2017:arXiv:}), if "magnetic" generalizations of the geodesic models of twin high frequency (HF) QPOs (\cite{Stu-Kot-Tor:2013:AAP:}) are applied to data. Observed HF QPOs and relativistic jets in microquasars indicate the presence of an external magnetic field, influencing both oscillations in the accretion disk and creation of jets. The presence of radiation-reaction force and its relevance in the vicinity of black holes can make sufficient contribution for the shifts in the QPO frequencies, which are usually observed. Moreover, the radiation reaction can support the accretion of charged particles from accretion disk towards the black hole. Depending on the magnitude of the external magnetic field, the radiation-reaction force can considerably shift the stable orbits of a particle which can sufficiently influence the predictions of black hole parameters.

The weakness of magnetic field in our study is understood in the sense that it does not perturb the spacetime metric; the corresponding condition on the field magnitude $B$ in the vicinity of Schwarzschild black hole of mass $M$ reads (\cite{Gal-Pet:1978:SovJETP:}):  
\beq \label{BBB}
B << B_{\rm G}={ \frac{c^4}{G^{3/2}} {M}_{\odot}}\left(\frac{{M}_{\odot}}{M}\right)\aprx10^{19}{ \frac{{M}_{\odot}} {M} }\,{\rm Gauss}\, .
\eeq
This weakness is compensated by large ratio $e/m$  for electrons and protons, whose motion will be essentially affected by magnetic fields already of the order of few gauss.

Study of the particle motion and electrodynamics in magnetized black holes began long ago, see e.g. \cite{Gal-Pet:1978:SovJETP:,Bla-Zna:1977:MNRAS:}, for a review of early results see \cite{Ali-Gal:1989A:SovPU:}. Equatorial orbits are of primary interest for theory of accretion disks. It was shown that the innermost stable circular orbits (ISCO) in the field of magnetized black hole are shifted towards the horizon (\cite{Gal-Pet:1978:SovJETP:}) for suitable direction of rotation.
Recently, detailed studies of the equatorial motion of charged particles in weakly magnetized Schwarzschild and Kerr black holes were performed in \cite{Fro-Sho:2010:PHYSR4:,Kol-Stu-Tur:2015:CQG:,Tur-Stu-Kol:2016:PHYSR4:}, revealing a number of new types of motion. 

Estimates show that in many physically interesting cases the radiation reaction force is non-negligible, and this has prompted us to consider what happens once the reaction force is taken into account. 
The general problem of the synchrotron radiation and treatment of the radiation-reaction force in curved spacetime was widely studied in literature. Below, we remind only few of them.
The electromagnetic radiation emitted by relativistic particles in the presence of strong gravity has been studied, e.g., in \cite{Poisson:2004:LRR:,Jon-Ruf-Zer:1973:PRL:,Zerilli:2000:NCBS:,Pri-Bel-Nic:2013:AmerJP:}. The problem of synchrotron radiation in curved spacetime has been studied, in \cite{Sokolov-etal:1983:SovPJ:,Sok-Gal-Pet:1978:PLA:} using covariantization of the flat space results. Actually, we will use here the same approach to describe the reaction force. For the recent treatment of radiation reaction in flat space see \cite{Gal-Spi:2006:GRC:} where the derivation of the Shott term was given explicitly along the lines of the Teitelboim idea to associate it with the Coulomb part of the electromagnetic field of a point charge. It is worth noting that the problem of the Shott contribution to reaction force (the third derivative term) still remains the source of discussion (see e.g. \cite{Poisson:2004:LRR:}), so we will touch it in this paper too. 

Generalization of radiation reaction equation to curved space was given by \cite{DeW-Bre:1960:AnnPhys:}. Their result (corrected by \cite{Hobbs:1968:AnnPhys:}) shows that radiation reaction in curved spacetime is essentially non-local and appeals to solve the integral equation. But contrary to gravitational radiation reaction in collisions of black holes, electromagnetic radiation from point particles can still be simplified to a local theory.

Recently, generalization of the radiation reaction equation to higher dimensions was discussed in \cite{Galtsov:2002:PRD:,Gal-Spi:2007:GrCosm:} which may be relevant to extra-dimensional models. Radiation from hypothetical massless charges was considered in \cite{Galtsov:2015:PRL:} bringing new light on the relationship between the ultrarelativistic and the massless limits in synchrotron radiation and radiation reaction equation. The motion of point particles with scalar, electric and mass charges, has been reviewed in \cite{Poisson:2004:LRR:}. The problem of radiation-reaction for extended charged particles has been studied in \cite{Cre-Tes:2011A:EPJP:}, its quantum aspects are discussed in \cite{Cre-Tes:2015A:EPJP:}. Statistical treatment of the radiation-reaction problem is given in \cite{Cre-Tes:2013A:PHYSR5:}. The problem of gravitational self-force of test particles is summarized in \cite{Barack:2014:book:}. Nevertheless, despite the active interest in the topic, and comprehensive literature review, the successful attempts to integrate the equations of motion in the curved background combined with external electromagnetic fields is quite rare.

In spite of weakness of radiation reaction force in ordinary stellar astrophysical situations, it was found recently that in high energy plasma processes in pulsar magnetospheres, black-hole accretion disks, hot accretion flows in X-ray Binaries and Active Galactic Nuclei, relativistic jets, and so on,  which are dominated by magnetic reconnection in plasma, radiation reaction may play a crucial role. The new theory, named radiative magnetic reconnection (for a recent review see \cite{Uzdensky:2016:MagRecAPSS:}), predicts notable radiative effects in astrophysical reconnection  such as radiation-reaction limits on particle acceleration, radiative cooling, radiative resistivity, braking of reconnection outflows by radiation drag, radiation pressure, viscosity, and even pair creation at highest energy densities.
The radiative reconnection theory is based on the flat-space description of radiation reaction adapted to phenomena in strong gravitational fields which is well justified in dense plasmas.

However, in the case of diluted media, when individual particle motion is relevant, it seems more adequate to use description of radiation reaction in curved space-time from the very beginning. This is the main goal of the present paper. Here, we would like both to discuss conceptual problems of such a description, and to analyse numerically the modification of particle motion within a simple model of weakly magnetized Schwarzschild black hole. Classifying the motion of charged particles by the given set of parameters, we find the explicit trajectories for each class of orbits and study the evolutions of the particle energy, angular momentum, etc.
 Motion of charged particles in the combined magnetic and gravitatational field without radiation reaction reveals interesting new types of trajectories. We intend to apply the full machinery of curved spacetime radiation reaction theory to these problems. We will also discuss possible applications of our results in some astrophysical scenarios.

The paper is organized as follows. In Sec.~\ref{sec-rr-flat}, we test the dynamical equations with radiation reaction force in flat spacetime using two main approaches and compare the results. In Sec.~\ref{Section-norad}, we present the properties of the circular motion of non-radiating charged particles in weakly magnetized Schwarzschild black hole, concentrating attention on the bounded orbits and charged particle oscillations. In Sec.~\ref{sec-rr-BH}, we give the general relativistic treatment of the radiation-reaction force in curved spacetime, and under reasonable assumptions we give explicit form of the equations of motion of charged particle in the vicinity of Schwarzschild black hole immersed in external asymptotically uniform magnetic field. In Sec.~\ref{section-traj}, we analyse the trajectories of charged particles for different classes of orbits and find the evolutions of the relevant parameters of the system during and after the decay of particle oscillations. Switching to the Gaussian units in Sec.~\ref{astro-section}, we estimate the characteristic decay times of the charged particle oscillations and discuss the relevance of the model in astrophysical situations. We summarize our results in Sec.~\ref{conclusion}. 

Throughout the paper, we use the spacetime signature $(-,+,+,+)$, and the system of geometric units in which $G = 1 = c$. However, for expressions having an astrophysical relevance, we use the constants explicitly. Greek indices are taken to run from 0 to 3.

\section{Radiation reaction in flat spacetime} \label{sec-rr-flat}
For completeness, we first summarize the results of the motion of charged particle in the flat spacetime. 
The equation which describes the motion of charged particle in magnetic field in general contains two forces
\beq \label{flateqmo1}
 \frac{\d u^\mu}{\d \tau} = f_L^\mu + f_R^\mu, 
\eeq
where $ f_L^\mu = (q/m) F^{\mu\nu} u_{\nu}$ is the Lorentz force, $F^{\mu\nu} = \partial^\mu A^\nu - \partial^\nu A^\mu$ is the tensor of electromagnetic field and $u^\mu (\tau) = dx^\mu/d\tau$ is a four-velocity of the particle. The last term $f_R^\mu$ is the radiation reaction force which in the non-relativistic case has to lead to the expression $\frac{3q^2}{2m} \frac{d^2 u^\alpha}{d \tau^2}$. Moreover any force vector has to satisfy the condition $f_R^\mu u_\mu = 0$. This implies that the correct covariant form of the expression for the radiation reaction force is 
\beq \label{flatradterm1}
f_R^\mu = \frac{2 q^2}{3 m} \left( \frac{d^2 u^\mu}{d\tau^2} + u^\mu u_\nu \frac{d^2 u^\nu}{d\tau^2} \right). \label{fradrec}
\eeq
This expression was found by Dirac and motion equation is sometimes called as Lorentz-Abraham-Dirac or Lorentz-Dirac (LD) equation. The first term in the parentheses, also known as the Schott term, arises from the particle electromagnetic momentum (\cite{Gal-Spi:2006:GRC:}). The second term in parentheses is the radiation recoil term which corresponds to the relativistic correction of the radiation reaction force. 
The four-velocity of the charged particle satisfies the following equations
\beq \label{norconflat}
u_\alpha u^{\alpha} = -1, \quad u_\alpha \dot{u}^{\alpha} = 0, \quad u_\alpha \ddot{u}^{\alpha} = - \dot{u}_{\alpha} \dot{u}^{\alpha}.
\eeq
{Thus, the motion of radiating charge is described by a third order differential equation in coordinates, rather than habitual second order.
This may lead to unphysical solutions - the existence of pre-acceleration in the absence of external forces. 
Even though the unphysical solutions can be removed by properly chosen initial conditions, the integration of exact form of equations of motion (\ref{flateqmo1}) and (\ref{fradrec}) is inconvenient due to the exponential increase of the computational error in practical calculations.
}

However, one can reduce the order of LD equation by the method proposed in \cite{Lan-Lif:1975:CTF:}, i.e., by rewriting the self-force in terms of the external force and the four-velocity of a particle. Substituting higher order terms in (\ref{fradrec}) by the derivatives of the Lorentz force we get the equation in the following form 
\beq \frac{\d u^\mu}{\d \tau} = f_L^\mu + \frac{2 q^2}{3 m} \left( \delta^{\mu}_{\alpha} + u^\mu u_\alpha \right) \frac{d f_L^\alpha}{d\tau}. \label{ladeq}
\eeq
This equation, usually referred as Landau-Lifshitz (LL) equation, has important consequences: it is of the second order, does not violate the principle of inertia, and the self-force vanishes in the absence of the external (Lorentz) force, \cite{Rohrlich:2001:PLA:,Poisson:1999:arXiv:}.
The self-contained derivation of the equation (\ref{ladeq}) in terms of retarded potentials is given in \cite{Poi-Pou-Veg:2011:LRR:}. The equation (\ref{ladeq}) can be applied for the cases with any external force acting on a charged particle instead of the Lorentz force. In case when $f_L^\mu = \frac{q}{m} F^{\mu\nu} u_{\nu}$, the radiation reaction force can be rewritten in the form
\beq 
f_R^\mu = k \q \left[F^{\mu\nu}_{\,\,\,\,, \alpha} u^{\alpha} u_{\nu} + \q (F^{\mu\nu} F_{\nu\rho} - F^{\beta\alpha} F_{\beta\rho} u_\alpha u^\mu) u^\rho \right], \label{radforceflat} 
\eeq
where $\q = q/m$ is the specific charge of the particle, $k = (2/3) \, \q \,q$,  and the comma in the first term denotes the partial derivative with respect to the coordinate $x^\alpha$. 
It was concluded in \cite{Spohn:2000:EURPHL:} that using the LL equation is identical to imposing Dirac's asymptotic condition $\lim_{\tau \rightarrow \infty} \dot{u}^{\mu} = 0$ to the LD equation. It was later confirmed by \cite{Rohrlich:2001:PLA:} that the reduced form of the equation of motion is exact, rather than approximative, though LL equation was proposed in \cite{Lan-Lif:1975:CTF:} as an approximative solution to the third order LD equation.
 More details on the treatment of radiation reaction of charged particles in flat spacetime can be found in a book of \cite{Spohn:2004:book:}. In our numerical study we found that the LL approximation is perfectly applicable if the Schott term is small with respect to the radiation recoil term, which is the case we consider here.
 Below we show the representative example of the charged particle motion in external uniform magnetic field, integrating both LD and LL equations. Results of numerical studies of LD and LL equations for the motion of charged particle in uniform magnetic field in flat spacetime are in accord with the analytical treatment of the radiation reaction force performed in \cite{Spohn:2000:EURPHL:}.

\subsection{Charged particle in uniform magnetic field} \label{subsec-rr-flat1}

\begin{figure*}
\includegraphics[width=\hsize]{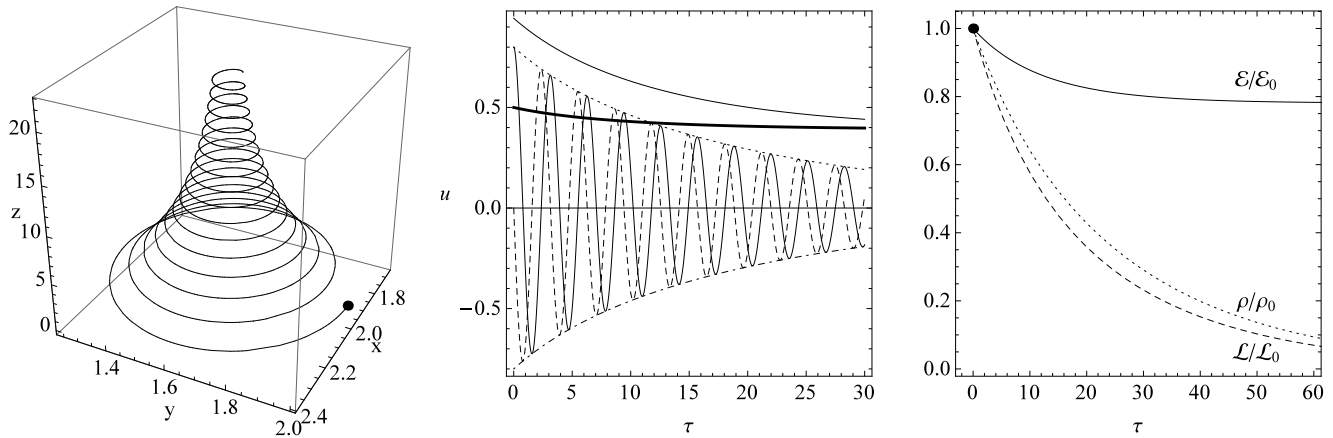}
\caption{
\label{trajflat} \label{flatELR}
Motion of radiating charged particle in flat spacetime. Left figure represents an example of 3D trajectory of charged particle. Middle plot represents the evolution of different components of 3-velocity in proper time $\tau$: the damping harmonic oscillations correspond to $u^x$ (solid thin) and $u^y$ (dashed) components of velocity, tangential to them (dotted) is a plane velocity of a particle $v_{\bot}$ orthogonal to $z$ axis, vertical $u^z$ component of velocity is shown by solid thick line and dot-dashed curve of the middle plot shows the evolution of 3-velocity ($\sqrt{u_x^2+u_y^2+u_z^2}$). Right figure shows the change of the energy, angular momentum and gyroradius of charged particle in time $t$ with respect to the static observer. Energy decreases up to the certain value (\ref{irreden}), while angular momentum and radius of gyration asymptotically tend to zero.
}
\end{figure*}

{
Let us consider the motion of a charged particle in a homogeneous magnetic field aligned along $z$-axis, such that the independent nonvanishing component of the electromagnetic tensor is $F_{x y} = B$. We introduce the new parameter in the form $\BB = q B/(2m)$, where the factor $1/2$ is added in order to correspond to the similar parameter introduced in the curved spacetime case.}

{ Both LD and LL equations, lead to equivalent results, differing in the number of initial conditions. In case of LD equation, one needs to set the values of $9$ constants - arbitrary independent components of initial position, velocity and the acceleration of the charged particle. The three other constants are given by the normalization condition (\ref{norconflat}). 
However, the direct integration of higher order equations leads to the exponential increase of the computational error in very short times. It is interesting to note that the problem of the time dispersion error can be greatly reduced by integrating equations of motion backwards in time. Similar method of solving Lorentz-Dirac equations has been proposed in the past by \cite{Hus-Bay:1976:PHYSR4:}.
}

On the other hand, the reduced-order equations of motion (\ref{ladeq}) can be written explicitly in the form
\bea 
\frac{d u^x}{d \tau } & = & 2 \BB u^y - 4 k \BB^2 \left(1 + u_{\bot}^2 \right) u^x, \label{flatux} \\
\frac{d u^y}{d \tau } & = & - 2\BB u^x - 4 k \BB^2 \left(1 + u_{\bot}^2 \right) u^y, \label{flatuy} \\
\frac{d u^z}{d \tau } & = & - 4 k \BB^2  u_{\bot}^2  u^z, \label{flatuz} \\ 
\frac{d u^t}{d \tau } & = & - 4 k \BB^2  u_{\bot}^2  u^t \label{flatut}.
\eea
Here $u_{\bot}^2  = (u^x)^2 + (u^y)^2$ is square of the particle velocity in the plane orthogonal to the magnetic field and $z$ axis. The representative trajectory of radiating charged particle corresponding to both LD and LL cases is shown in Fig.\ref{trajflat}. Initial conditions are chosen as follows: initial plane velocity is $u_{\bot 0} = 0.8 c$, velocity in vertical direction is $u^z_0 = 0.5 c$, magnetic field is aligned along $z$ axis and magnetic parameter is chosen to be $\BB = 1$, radiation parameter $k=0.01$. The estimations of the parameters in realistic situations are given in Sec~\ref{astro-section}. Energy and angular momentum loss leads to decay of the plane velocity of the particle, while the vertical component remains constant. Apparent deceleration along $z$ axis shown in Fig.\ref{trajflat} (solid thick curve in the middle plot), while no forces are applied in the $z$ direction, appears only in the frame moving with the particle. The velocity with respect to the static observer, $v^z = dz/dt \equiv u^z/u^t = const$, remains constant.


\subsection{Energy and momentum loss} \label{subsec-E-L-loss-flat}

The rate of the energy loss of the particle can be evaluated from Eq.(\ref{flatut}). Modifying it for a static observer, we get
\beq
\frac{d E}{d t}  =  - k B^2 \q^2  u_{\bot}(t)^2 . \label{flatEt} 
\eeq
Integrating this equation, one can obtain the energy of the particle in a given moment of time. The evolution of the energy in time is shown in Fig.\ref{flatELR} for the given trajectory. Thus the energy loss will be given only by the change of the plane velocity $u_{\bot}$ in time, while the kinetic energy associated with the motion in the $z$-direction will be conserved. The plane velocity $u_{\bot}$ of the particle decreases in time and asymptotically tends to zero as represented by the dotted curve of the middle plot of Fig.\ref{trajflat}.
 This implies that there exists an irreducible (specific) energy of radiating charged particle which corresponds to the final state of the particle, having the following simple form
\beq \label{irreden}
\ce_{0} = \left(1 - (v_0^z)^2\right)^{-\frac{1}{2}},
\eeq
where $v_0^z$ is the vertical velocity of the particle along the $z$-axis, measured by the static observers, which is constant during the radiation process.
  
In order to find the rate of the angular momentum loss, one can fix the motion in a plane by taking $u^z = 0$.
In general, the specific angular momentum for the motion in a plane is defined by the formula $\cl = \rho^2 {d \phi}/{d\tau} + \q A_{\phi}$, where $\rho$ is gyroradius of the particle trajectory. Reminding that ${d \phi}/{d\tau} = u_{\bot} \gamma / \rho$ and $\rho = u_{\bot}/\omega_L$, where $\omega_L = q B/m \equiv 2\BB$ is the Larmor frequency, one can write the specific angular momentum of the radiating charged particle in given moment of time in the form
\beq \cl = \frac{u_{\bot}(\tau)^2}{4 \BB} \left[2\gamma(\tau) + 1\right],  \quad \gamma = (1-u_{\bot}^2)^{-\frac12}. \label{angmomflat} \eeq
Solving first two equations of motion (\ref{flatux}) and (\ref{flatuy}), and substituting into Eq.(\ref{angmomflat}), we get the evolution of the angular momentum in time, which is represented in Fig.\ref{flatELR}. Unlike the energy of the particle, the angular momentum asymptotically tends to zero for large $\tau$. This occurs due to the reason that the gyroradius $\rho$ of the charged particle tends to zero as well, while $\cl$ is proportional to $\rho$.

One can find the ratio between the angular momentum loss $\dot{\cl} = d \cl/dt$ and the energy loss $\dot{\ce} = d \ce/dt$ in the form 
\beq
\frac{\dot{\cl}}{\dot{\ce}} = r^2 \Omega \frac{u_{\bot}^2 + 1}{u_{\bot}^2}. \label{dldeflat}
\eeq
where $\Omega$ is the angular frequency of the charged particle measured by the observers at rest. In Cartesian coordinates $\Omega$ takes the form
\beq \Omega = \frac{x \dot{y} - y \dot{x}}{x^2+y^2},
\eeq
where dots denote the derivative with respect to the coordinate time $t$.

\subsection{Decay time} \label{subsec-rr-flat2}

One can find the characteristic time required to decay the energy of a radiating charged particle in the following way. Since the velocity in the magnetic field direction is constant, one can consider only the planar motion of the particle by taking $u^z=0$. This implies that according to condition $u^\alpha u_\alpha = -1$, we get $u_{\bot}^2 = (u^t)^2 - 1$. Thus, equation (\ref{flatut}) can be rewritten in the form
\beq
\frac{d \ce}{d \tau} = - \KK \left(\ce^3 - \ce\right), \quad  \KK = 4 k \BB^2.
\eeq
Integrating this equation, we get the particle energy in a given moment of time
\beq \label{ener-flat-tau}
\ce(\tau) = \frac{\ce_i e^{\KK \tau}}{\sqrt{1 + \ce_i^2 \left(e^{2\KK\tau} - 1 \right)}},
\eeq
where the integration constant $\ce_i$ is the initial specific energy of the particle. Asymptotically in time the specific energy tends to the particle rest energy, being equal to 1. Thus, the decay time during which the specific energy will be lowered from $\ce_i$ to $\ce_f$ due to radiation takes the following form
\beq \label{cooltimeflat1}
{\cal T} = \frac{1}{2 \KK} \ln{\frac{\ce_f^2 (\ce_i^2 - 1)}{\ce_i^2 (\ce_f^2 - 1)}}.
\eeq
where $\KK = 4 k \BB^2$. Since the energy in (\ref{ener-flat-tau}) is the exponentially decreasing function, the particle energy cannot be reduced to $1$ in practical calculations.

\section{Charged particle orbiting black hole without radiation-reaction} \label{Section-norad}

In this section we shortly summarize previous results related to the particle motion in the field of magnetized black holes presented in \cite{Kol-Stu-Tur:2015:CQG:,Stu-Kol:2016:EPJC:,Tur-Stu-Kol:2016:PHYSR4:}.
The interval in the \Schw{} black hole spacetime in sperical coordinates ($t,r,\theta,\phi$) reads 
\beq
 d s^2 = -f(r) d t^2 + f^{-1}(r) d r^2 + r^2(d \theta^2 + \sin^2\theta d \phi^2), \label{SCHmetric}
\eeq
where $M$ is the black hole mass and the function $f(r)$ is the lapse function given by
\beq \label{lapsefun}
		f(r) = 1 - \frac{2 M}{r}. 
\eeq
Hereafter, without loss of generality we put the mass of the black hole to be equal to unity, $M=1$.
Let us consider the black hole immersed into external asymptotically uniform magnetic field. In the case when this field is weak implying that the metric of Schwarzschild black hole is not violated (see Eq.(\ref{BBB})), one can write the solution of the Maxwell equation for the four-vector potential of electromagnetic field $A^\mu$ in the following form (\cite{Wald:1984:book:})
\beq
A_{\phi} = \frac{B}{2} \, g_{\phi\phi}  = \frac{B}{2} \, r^2 \sin^2 \theta, \label{aasbx}
\eeq
which is the only nonzero component of four-vector potential $A^\mu$ and $B$ is the strength of the magnetic field at spatial infinity, which is taken to be constant. The antisymmetric tensor of the electromagnetic field $F_{\mu \nu} = A_{\nu,\mu} - A_{\mu,\nu}$ in this case has only two independent nonzero components
\beq \label{FaradayUniform}
F_{r \phi} = B r \sin^2{\theta}, \quad F_{\theta \phi} = B r^2 \sin{\theta} \cos{\theta}.
\eeq
%

\subsection{Bounded motion around black hole}

\begin{figure*}
\includegraphics[width=\hsize]{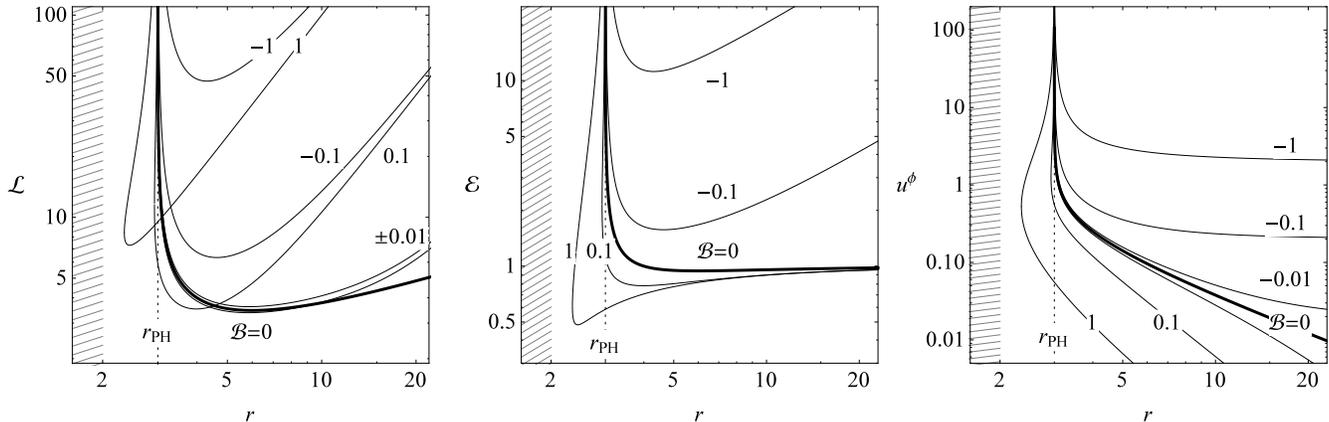}
\caption{\label{trajrad0}
Dependence of angular momentum $\cl$, energy $\ce$ and azimuthal velocity $u^\phi$ of a charged particle on the location of the circular orbit $r$ without radiation for different values of magnetic parameter $\BB$.
}
\end{figure*}

The motion of charged particles is described by the Lorentz equation in curved spacetime. In case when the radiative processes can be neglected, one can write for the particle of mass $m$ and charge $q$ the Lorentz equation of motion
\beq \label{eqmonorad}
 \frac{D u^\mu}{d \tau} \equiv \frac{\d u^\mu}{\d \tau} + \Gamma^\mu_{\alpha\beta} u^\alpha u^\beta = \frac{q}{m} F^{\mu}_{\,\,\, \nu} u^{\nu}, 
\eeq
where  $u^{\mu}$ is the four-velocity of the particle, normalized by the condition $u^{\mu} u_{\mu} = - 1$, $\tau$ is the proper time of the particle and components of $\Gamma^\mu_{\alpha\beta}$ are the Christoffel symbols.

Symmetry of the Schwarzschild black hole (\ref{SCHmetric}) and external magnetic field give us a right to find the conserved quantities associated with the time and space components of the generalized four-momentum $\cp_\alpha = p_\alpha + A_\alpha$. Thus, the energy and angular momentum of the charged particle in the presence of external uniform magnetic field take the following form
\bea
 E &=& - \cp_t = m f(r) \frac{\d t}{\d \tau}, \\
 L &=& \cp_\phi = m r^2 \sin^2\theta \left(\frac{\d \phi}{\d \tau} + \frac{q B}{2m} \right). \label{angmom}
\eea
In this case, the circular motion of charged particles is always bounded in the plane orthogonal to the magnetic field lines, which corresponds to the equatorial plane $\theta = \pi/2$. The boundary of the motion is governed by the shape of the effective potential, determined by the equation
\beq
 \ce^2 = V_{\rm eff} (r,\theta; \cl,\cb). \label{MotLim}
\eeq
Using the notation 
\beq
\ce = \frac{E}{m}, \quad \cl = \frac{L}{m}, \quad \cb = \frac{q B}{2m}.
\eeq
we can write the effective potential in the form
\beq 
V_{\rm eff} (r,\theta) \equiv f(r) \left[1+\left(\frac{\cl}{r \sin{\theta} } - \cb\, r \sin{\theta}\right)^2\right]. \label{VeffCharged} 
\eeq
The effective potential (\ref{VeffCharged}) shows clear symmetry $(\cl,\cb)\leftrightarrow(-\cl,-\cb)$ that allows to distinguish the following two situations 
\begin{itemize}
\item[-] {\it minus configuration}, with $\cl>0, \cb<0$  (equivalent to $\cl<0, \cb>0$) - magnetic field and angular momentum parameters have opposite signs and the Lorentz force is attracting the charge towards the black hole.
\item[+] {\it plus configuration}, with $\cl>0, \cb>0$ (equivalent to $\cl<0, \cb<0$) - magnetic field and angular momentum parameters have the same signs and the Lorentz force is repulsive, acting outward the black hole.
\end{itemize}
The last case appears only in the combination of gravity with electromagnetic field and cannot exist in flat spacetime
(for details, see \cite{Kol-Stu-Tur:2015:CQG:,Tur-Stu-Kol:2016:PHYSR4:,Stu-Kol:2016:EPJC:}). Particular trajectories of charged particles in magnetized Schwarzschild black hole are compared with radiating particle motion in Section~\ref{section-traj}.

\subsection{Charged particle oscillations}


Charged particles can undergo stable quasi-harmonic oscillations in radial and vertical directions in the vicinity of magnetized black hole. 
In the case when the oscillations are small enough in comparison with the radius of the corresponding stable circular orbit, one can find the locally measured frequencies of vertical $\omega_{\mit}$ and radial $\omega_{\mir}$ oscillations equal to
\bea
\omega^2_{\mit} &=& \frac{\cl_c^2}{r^4}-\cb^2, \\
\omega^2_{\mir} &=& \frac{1}{(r-2) r^5} \big[ (r-2)^2 \left(\cb^2 r^4+3 \cl_c^2\right) \nonumber \\ 
&& \qquad\qquad\quad -2r \left( \cb r^2 -\cl_c \right)^2 -2r^3  \big]
\eea
where $\cl_c$ is the specific angular momentum at the circular orbit, for details see \cite{Kol-Stu-Tur:2015:CQG:}. In addition to the frequencies given above, one can find also the Keplerian axial frequency $\omega_\mip$ and Larmor angular frequency $\omega_{\mil}$ which are given by
\beq \label{freq-omK-omB}
 \omega_\mip = \frac{\d \phi}{\d \tau} = U^\phi = \frac{\cl_c}{g_{\phi\phi}} - \cb, \quad \omega_{\mil} = \frac{q B}{m} = 2 \cb.
\eeq
As one can see, the frequency $\omega_{\mil}$ does not depend on $r$ coordinate and plays crucial role in the regions detached from the black hole. The characteristic oscillations of charged particles and representative plots of charged particle trajectories around Schwarzschild black hole can be found in Fig.7 of paper \cite{Kol-Stu-Tur:2015:CQG:}. For the particle to oscillate in the vicinity of a black hole, its energy has to be larger than the minimum of the effective potential, however, keeping the finite type of the motion. Thus, the minimal energy of a charged particle in trapped states corresponds to the stable circular orbits given by the minimum of the effective potential (\ref{VeffCharged}). The maximal energy at the trapped states is given by the unstable circular orbit with the corresponding angular momentum. Energy of the charged particle at the circular orbit is given by
\beq \label{trappedenergy}
\ce_{\pm}^2 = \frac{r f^2 }{(r-3)^2} \left(r-3+2 \BB^2 r^3 f \pm 2 \BB r \sqrt{r-3+ \BB^2 r^4 f^2} \right),
\eeq
where the signs correspond to the maximal and minimal energy of a particle at the circular orbit, and $f = 1-2/r$ is the lapse function. The difference between $\ce_{+}$ and $\ce_{-}$ can be very large for large values of magnetic parameter $\BB$, representing the charged particles accelerated up to ultrarelativistic velocities (\cite{Kol-Stu-Tur:2015:CQG:,Stu-Kol:2016:EPJC:}).

Some types of the quasi-circular epicyclic motion are possible only in the presence of magnetic field. One of the interesting and illustrative examples of the effect of magnetic field on the charge particle motion in the black hole background is the appearance of curled trajectories, as demonstrated in Fig.\ref{trajrad2} which corresponds to the {\it plus configuration} with repulsive Lorentz force. 
The conservation of the angular momentum (\ref{angmom})  in the absence of radiation leads to the equation
\beq
 \dot{\phi} = \frac{\cl}{r^2} - \cb.
\eeq
In the non-magnetic case, the right hand side of the equation above is positive for positive $\cl$. The presence of magnetic field with $\cb>0$ decreases the velocity in $\phi$-direction that can become negative, if
\beq \label{curl-L}
 \cl > \cl_{*}(r;\cb) \equiv \cb r^2.
\eeq
For the energy, this condition means
\beq \label{enschCO}
 \ce > \ce_{*}(\cl;\cb) \equiv \sqrt{1- 2\sqrt{\cb/\cl}}.
\eeq
Thus, decreasing the azimuthal velocity of a particle with the influence of magnetic field, and keeping the above given conditions, we get so called curled trajectories, which are not possible in the absence of magnetic field. This type of orbits does not appear in the case of attractive Lorentz force corresponding to the {\it minus configurations}.

In the next sections we show that the oscillations of charged particles in magnetic field near a Schwarzschild black hole will be damped due to the synchrotron radiation and, in particular cases the charged particle orbits will not be able to stay stable, bringing the particle to the black hole due to the effect of the radiation reaction force.

\section{Radiation reaction in the field of magnetized black holes} \label{sec-rr-BH}

\subsection{Radiation-reaction force}

The motion of a relativistic charged particle is governed by the Lorentz-Dirac equation which includes the influence of the external electromagnetic fields and corresponding radiation-reaction force. The last force arises from the radiative field of the charged particle and the equations of motion in general can be written in the form
\beq \label{cureqmogen1}
   \frac{D u^\mu}{\d \tau} = \q F^{\mu}_{\,\,\,\nu} u^{\nu} + \q {\cal F}^{\mu}_{\,\,\,\nu}  u^{\nu}, 
\eeq
where the first term on the right hand side of Eq.(\ref{cureqmogen1}) corresponds to the Lorentz force with electromagnetic tensor $F_{\mu\nu}$ given by (\ref{FaradayUniform}), while the second term is the self-force of charged particle with the radiative field ${\cal F}_{\mu\nu} = {\cal A}_{\nu,\mu} - {\cal A}_{\mu,\nu}$. The vector potential of the self-electromagnetic field of the charged particle satisfies the wave equation 
\beq \label{wave-ret-pot}
\Box {\cal A}^{\mu} - R^{\mu}_{\,\,\,\nu} {\cal A}^{\nu} = - 4 \pi j^{\mu},
\eeq
where $\Box = g^{\mu\nu} D_{\mu} D_{\nu}$, and $D_{\mu}$ is the covariant differentiation and $R^{\mu}_{\,\,\,\nu}$ is the Ricci tensor. The retarded solution to Eq.(\ref{wave-ret-pot}) for the vector potential takes the form
\beq
{\cal A}^{\mu} (x) = q \int G_{+\lambda}^{\mu} (x,z(\tau)) u^{\lambda} d\tau,
\eeq
where $G_{+\lambda}^{\mu}$ is the retarded Green function and the integration is taken along the worldline of the particle $z$, i.e., $u^{\mu} ( \tau) = d z^{\mu} ( \tau ) / d \tau$. For details, see, e.g. \cite{Poisson:2004:LRR:}. The covariant generalization of the dynamics of radiating charged particle in curved spacetime has been derived in \cite{DeW-Bre:1960:AnnPhys:} and completed in \cite{Hobbs:1968:AnnPhys:}, using the tetrad formalism. The explicit form of Eq.(\ref{cureqmogen1}) for the motion of charged particle undergoing radiation-reaction force in curved spacetime reads
\bea 
&& \frac{D u^\mu}{\d \tau} = \q F^{\mu}_{\,\,\,\nu} u^{\nu} 
+ \frac{2 q^2}{3 m} \left( \frac{D^2 u^\mu}{d\tau^2} + u^\mu u_\nu \frac{D^2 u^\nu}{d\tau^2} \right) \nonumber \\ 
&& + \frac{q^2}{3 m} \left(R^{\mu}_{\,\,\,\lambda} u^{\lambda} + R^{\nu}_{\,\,\,\lambda} u_{\nu} u^{\lambda} u^{\mu} \right) + \frac{2 q^2}{m} ~f^{\mu \nu}_{\rm \, tail} \,\, u_\nu, 
\label{eqmoDWBH}  
\eea 
where the last term of Eq.(\ref{eqmoDWBH}) is the tail integral
\beq
f^{\mu \nu}_{\rm \, tail}  = \int_{-\infty}^{\tau-0^+}     
D^{[\mu} G^{\nu]}_{ + \lambda'} \bigl(z(\tau),z(\tau')\bigr)   
u^{\lambda'} \, d\tau' .
\eeq
Detailed derivation of the equations of motion of radiating charged particles can be found in \cite{Hobbs:1968:AnnPhys:,Poisson:2004:LRR:}. The integral in the tail term is evaluated over the past history of the charged particle with primes indicating its prior positions. All other quantities are evaluated at the current position of the particle $z(\tau)$. The term containing the Ricci tensor vanishes in the vacuum metrics, so this term is irrelevant in our case. The existence of the "tail" integral in (\ref{eqmoDWBH}) implies that the radiation reaction in curved spacetime has non-local nature, because the motion of the charged particle depends on its whole history and not only on its current state. The radiation field ${\cal F}_{\mu\nu}$ in Eq.(\ref{cureqmogen1}), emitted by the charged particle, interacts with the curvature of background spacetime and comes back to the particle with a delay corresponding to the tail integral in (\ref{eqmoDWBH}). In such sense, the radiated electromagnetic field of charged particle carries the information about the history of the particle. Even in the absence of external forces, such as the Lorentz force, the free trajectory of the charged particle does not follow the geodesics, which is one of the most important consequences of equation (\ref{eqmoDWBH}).

However, for purposes of the present paper, the tail term can be neglected as we show below. The tail terms can be estimated based on the results by \cite{Dew-Dew:1964:PHYSNY:}, \cite{Smi-Wil:1980:PRD:}, as well as multiple subsequent papers.
For a particle with the charge $q$ and mass $m$, the ratio of the tail force $F_{\rm tail} \sim G M q^2/(r^3 c^2)$  to the Newton-force $F_{\rm N} \sim  G M m/r^2$
in the vicinity of a black hole ($r\sim r_H = 2 G M/c^2$) of the stellar mass $M \sim 10 M_{\odot}$ is
\beq
         \frac{F_{\rm tail}}{F_{\rm N}}   \sim     \frac{q^2}{m M G}   \sim   10^{-19} \left(\frac{q}{e}\right)^2 \left(\frac{m_e}{m}\right) \left(\frac{10 M_{\odot}}{M}\right),
\eeq
where $e$ and $m_e$ are the charge and the mass of an electron. For supermassive black holes (SMBH) with the mass $M \sim 10^9 M_{\odot}$ this ratio is $8$ orders lower.
On the other hand, the radiation reaction force (second term on the right hand side of (\ref{eqmoDWBH})) depends on the presence of external force, arising in our case from the external magnetic field. According to \cite{Pio-etal:2011:ASBULL:,Baczko-etal:2016:AAP:}, the characteristic values of the magnetic fields near the stellar mass black holes and SMBH are $B \sim 10^8 {\rm G}$, for $M = 10 M_{\odot}$ and $B \sim 10^4 {\rm G}$, for $M = 10^9 M_{\odot}$. Thus, for the particle with velocity comparable to the speed of light, $v\sim c$, the ratio of the radiation reaction force $F_{\rm RR} \sim q^4 B^2/(m^2 c^4)$ to the Newton-force gives an order
\beq \label{FRRtoFN}
 \frac{F_{\rm RR}}{F_{\rm N}}   \sim \frac{q^4 B^2 M G}{m^3 c^8}  \sim  10^{3} \left(\frac{q}{e}\right)^4 \left(\frac{m_e}{m}\right)^3 \left( \frac{B}{10^8 {\rm G}}\right)^2 \left(\frac{M}{10 M_{\odot}}\right).
\eeq
The ratio of $F_{\rm RR}$ to $F_{\rm N}$ for SMBH with $M = 10^9 M_{\odot}$ and magnetic field $B \sim 10^4 {\rm G}$ gives the same order of magnitude.

{The above estimations of the tail term apply in the non-relativistic, or moderately relativistic case when the Lorentz factor is of the order of unity. More general argument is based on the treatment of the gravitational self-force in the Schwarzschild and Kerr spacetimes. As was shown by one of the present authors in \cite{Galtsov:1982hwm}, the radiative part of the self-force (based on the half-difference of the retarded and advanced Green's function) in the Kerr metric satisfies the (averaged on time) balance equations for the energy and angular momentum of radiation of all spins $s=0,\,1,\,2$ in the sense that local work of the self-force is equal to radiated fluxes at infinity and the black hole horizon. This is valid independently of the particle velocity and extends to the ultrarelativistic case. Later on it was shown that similar balance is valid for the Carter's constant (\cite{Mino:2003yg}), which completes the set of quantities determining geodesic motion in the Kerr field. The conservative part of the self-force (half-sum of the retarded and advanced potentials) is more difficult to compute since this demands proper elimination of divergences. This caused a vivid discussion in the literature from the mid-90-ies to early 2000-ies, nicely reviewed in \cite{Tanaka:2005ue} (for more recent work on the same subject containing further references see \cite{Fujita:2016igj}), resulting in a consensus opinion that this contribution is small with respect to to radiative self-force especially with growing Lorentz factor (\cite{Pound:2005fs,Sago:2005gd}). Therefore, if one is interested to estimate the gravitational and electromagnetic tail terms in the ultrarelativistic limit, using the above results one can revoke the gravitational synchrotron radiation (GSR), which was computed earlier for spins $s=0,\,1,\,2$ most notably in \cite{Chrzanowski:1974nr} and in \cite{Breuer1975} with comparison to flat-space synchrotron radiation (SR). From these results one can see that GSR in the ultrarelativistinc limit is suppressed by a square of the Lorentz factor with respect to SR.
These arguments lead us to conclude that for elementary particles moving with any velocity in both gravitational and electromagnetic field the purely gravitational tail term can be neglected in comparison with the electromagnetic (properly covariantized) radiation reaction force.
}
Thus, for purposes of the present paper, the equation of motion (\ref{eqmoDWBH}) can be simplified to the following covariant form of LD equation:
\beq \label{cureqmo1}
   \frac{D u^\mu}{\d \tau} = \q F^{\mu}_{\,\,\,\nu} u^{\nu} + f_R^\mu,
\eeq
with the radiation reaction force given by
\beq f_R^\mu = \frac{2 q^2}{3 m} \left( \frac{D^2 u^\mu}{d\tau^2} + u^\mu u_\nu \frac{D^2 u^\nu}{d\tau^2} \right). \label{curfradrec}
\eeq
Introducing the four-acceleration as a covariant derivative of four-velocity, $a^{\mu} = D u^{\mu} / d\tau$, one can rewrite the term $D^2 u^{\mu} / d\tau^2$ as follows
\bea 
&&\frac{D^2 u^{\mu}}{d\tau^2} \equiv \frac{D a^\mu}{d \tau} = \frac{d a^\mu}{d\tau} + \Gamma^{\mu}_{\alpha\beta} u^\alpha a^\beta 
\nonumber \\
&&= \frac{d}{d\tau} \left(\frac{d u^\mu}{d\tau} + \Gamma^{\mu}_{\alpha\beta} u^\alpha u^\beta \right) + \Gamma^{\mu}_{\alpha\beta} u^\alpha \left(\frac{d u^\beta}{d\tau} + \Gamma^{\beta}_{\rho\sigma} u^\rho u^\sigma \right) 
\nonumber \\
&&= \frac{d^2 u^\mu}{d \tau^2} + \left( \frac{\partial \Gamma^{\mu}_{\alpha\beta}}{\partial x^\gamma} u^\gamma u^\beta + 3 \Gamma^{\mu}_{\alpha\beta}  \frac{d u^\beta}{d\tau} + \Gamma^{\mu}_{\alpha\beta} \Gamma^{\beta}_{\rho\sigma} u^\rho u^\sigma \right) u^\alpha. \nonumber \\
  \label{curdadtau}
\eea
Thus, the general relativistic equations of radiating charged particle motion are given by Eqs (\ref{curfradrec}) and (\ref{curdadtau}). However, the full form of the equations is plagued by runaway solutions. One can avoid this problem in a similar way, as in the flat spacetime case, namely reducing the order of differential equations. In the absence of the radiation-reaction force, the motion of the charged particle in external electromagnetic field is governed by equation (\ref{eqmonorad}). Taking the covariant derivative with respect to the proper time from both sides of Eq.(\ref{eqmonorad}), we get
\beq \label{curdadtauFab}
\frac{D^2 u^{\alpha}}{d\tau^2} = \q \frac{D F^{\alpha}_{\,\,\,\beta}}{d x^{\mu}} u^\beta u^\mu + \q^2 F^{\alpha}_{\,\,\,\beta} 
F^{\beta}_{\,\,\,\mu} u^\mu, 
\eeq
Substituting (\ref{curdadtauFab}) into Eq.(\ref{curfradrec}), we get the radiation reaction force in the form
\beq \label{curradforce}
f_R^\alpha = k \q \left(\frac{D F^{\alpha}_{\,\,\,\beta}}{d x^{\mu}} u^\beta u^\mu + \q \left( F^{\alpha}_{\,\,\,\beta} 
F^{\beta}_{\,\,\,\mu} +  F_{\mu\nu} F^{\nu}_{\,\,\,\sigma} u^\sigma u^\alpha \right) u^\mu \right),
\eeq
where the covariant derivative from the second rank tensor reads
\begin{equation}
\frac{D F^{\alpha}_{\,\,\,\beta}}{d x^{\mu}} = \frac{\partial F^{\alpha}_{\,\,\,\beta}}{\partial x^{\mu}} + \Gamma^{\alpha}_{\mu\nu} F^{\nu}_{\,\,\,\beta} - 
\Gamma^{\nu}_{\beta\mu} F^{\alpha}_{\,\,\,\nu}.
\end{equation}
\\
Equations (\ref{cureqmo1}) and (\ref{curradforce}) give a covariant form of Landau-Lifshitz equations. Below we test these equations in particular case of the motion of charged particles around Schwarzschild black hole immersed into external asymptotically uniform magnetic field.

The motion of charged particles in the vicinity of magnetized 
Schwarzschild black hole is generally chaotic \cite{Kop-Kar:2014:APJ:,Kop-etal:2010:APJ:}.
However, close to the minimum of the effective potential, which corresponds to the stable circular orbit, the motion of the charged particle is regular, being of harmonic character.
The motion is also regular, if the particle is moving entirely in the equatorial plane and the chaotic behaviour appears with increasing the inclination angle.
Here, we focus our attention on the regular motion only. In order to represent the equations of motion explicitly, we fix the plane of the motion at the equatorial plane, $\theta = \pi/2$, of a magnetized black hole. However, for exploring the trajectories of particles, we use the full set of equations of motion and solve them numerically. Without loss of generality, one can again equalize the mass of a black hole to unity, $M=1$. The non-vanishing components of equations of motion of radiating charged particles moving around Schwarzschild black hole immersed into external asymptotically uniform magnetic field take the following form
\begin{widetext}
\bea
\frac{d u^t}{d \tau } = \frac{2 u^t u^r }{r (2-r)} -  \frac{2 k \mathcal{B} u^t}{r} \left\{2 \mathcal{B} r f \left[f (u^{t})^2 - 1 \right] - u^{\phi} \right\}, \label{radeqSch-ut} \\
\frac{d u^r}{d \tau } = u^{\phi} (2 \BB r f + r - 1) - \frac{1}{r^2}
- \frac{2 k \mathcal{B} u^r}{r}  \left\{2 \mathcal{B} r f^2 (u^t)^2 - u^{\phi} \right\}, \\
\frac{d u^{\phi }}{d\tau }= - \frac{2 u^r \left(u^{\phi} + \mathcal{B} \right)}{r} 
+ \frac{2 k \mathcal{B}}{r^3} \left[r^2 (u^{\phi})^2 + 2 \BB r^3 f^2 (u^{t})^2 u^{\phi} + 1 \right],
\eea
\end{widetext}
where $f$ is the lapse function given by (\ref{lapsefun}), $\BB = q B/(2 m)$ is the magnetic parameter and $k = 2 q^2/(3 m)$ is the radiation parameter. 

\subsection{Energy and angular momentum loss}

The total energy-momentum radiated by the charged particle is equal to the integral of the radiation reaction force taken along the worldline of the particle. In flat spacetime, the radiated four momentum of a particle with charge $q$ is given by $d P^{\mu}/d\tau = \frac{2}{3}  q^2 a^{\alpha} a_{\alpha} u^{\mu}$.
Synchrotron radiation in curved spacetime has been studied, in \cite{Sokolov-etal:1983:SovPJ:}, \cite{Sok-Gal-Pet:1978:PLA:} using covariantization of the flat space results. The problem has been revisited more recently in \cite{Shoom:2015:PHYSR4:}, where, however, the radiation reaction force is not taken into account. The evolutions of
the particular components of the four-momentum of the particle with radiation reaction force can be found from equations (\ref{cureqmo1}) and (\ref{curradforce}).
For the motion at the equatorial plane, the energy loss is given by
\beq \label{enerlossSCH}
\frac{d {\cal E}}{d\tau} = - 2 k \BB \left[ 2 \BB \ce^3 - \ce \left(2\BB f + \frac{u^\phi}{r}\right) \right].
\eeq
For ultrarelativistic particle with $\ce \gg 1$, the leading contribution to the energy loss is given by the first term in square brackets of (\ref{enerlossSCH}). However, for small velocities close to the stable circular orbits, the last two terms can play significant role. Both situations will be studied in details in the following section. 
Similarly, one can find the rate of angular momentum loss as
\beq  \label{angmomlossSCH}
\frac{d {\cal L}}{d\tau} = 4 \BB^2 k u^{\phi} \left( f^2 (u^t)^2  - f \right) - 2 u^r u^{\phi} \left(r - 4 \BB^2 k^2 \right) + 2 r \BB u^r.
\eeq
In the next section we will test the equations of motion numerically for general bounded motion of radiating charged particle without restriction to the equatorial plane.

\begin{figure*}
\includegraphics[width=\hsize]{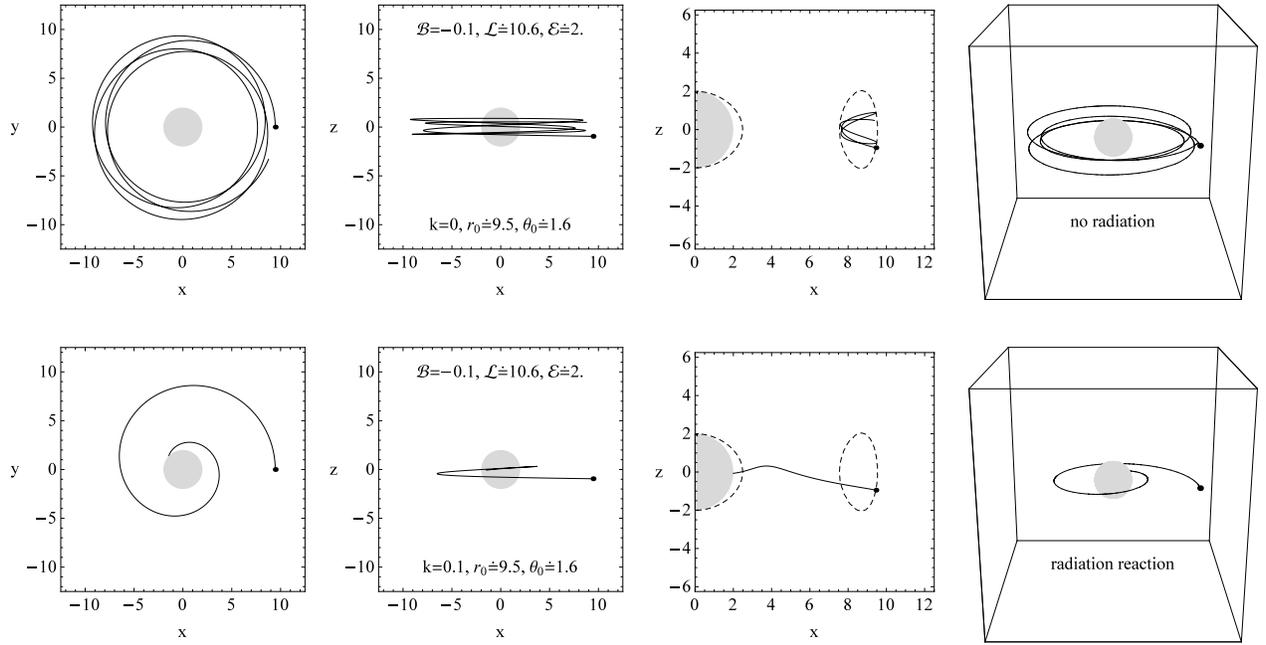}
\caption{\label{trajrad1}
Representative comparison of trajectories of non-radiating (first line) and radiating (second line) charged particles with {\it minus configuration} corresponding to the attractive Lorentz force from different view angles around black hole in magnetic field. Initial conditions in both cases are chosen to be same and shown in the second row plots. Starting point is indicated by black dot. A radiating charged particle escapes the initial boundary of the motion (dashed contours in the third row) governed by an effective potential (\ref{VeffCharged}) and collapses to the black hole due to the loss of angular-momentum. Magnetic field is aligned with $z$ - axis.
Trajectory of radiating charged particle corresponds to the integration of full set of equations of motion (\ref{cureqmo1}) and (\ref{curradforce}) without fixing the plane of the motion. 
}
\end{figure*}
\begin{figure*}
\includegraphics[width=\hsize]{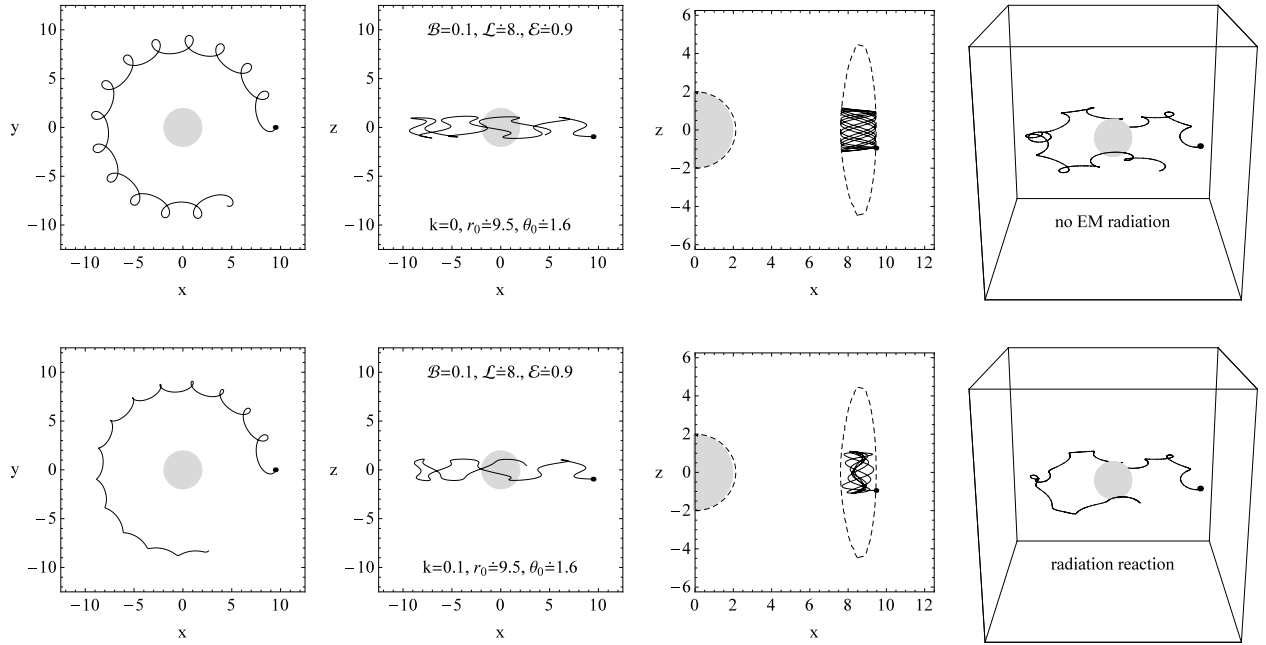}
\caption{\label{trajrad2}
Similar to Fig.\ref{trajrad1} comparison of trajectories of non-radiating and radiating charged particles in case of repulsive Lorentz force, i.e. {\it plus configuration}. 
 A radiating particle stays in the region of initial boundary of the motion and radiates its oscillatory part of energy-momentum tending to the stable circular orbit.
}
\end{figure*}

\section{Damping of oscillations} \label{section-traj}

In this section we study the damping of charged particle oscillations due to radiation reaction force for both $"+"$ and $"-"$ configurations. In particular, we demonstrate that the particles at $"-"$ configurations spiral down to the black hole, while at $"+"$ configurations the motion remains stable. In the subsection 5.3, we study the evolution of circular orbits under the influence of self force. We show that in such a case the circular orbits can be shifted outwards from the black hole.

\begin{figure*}
\includegraphics[width=\hsize]{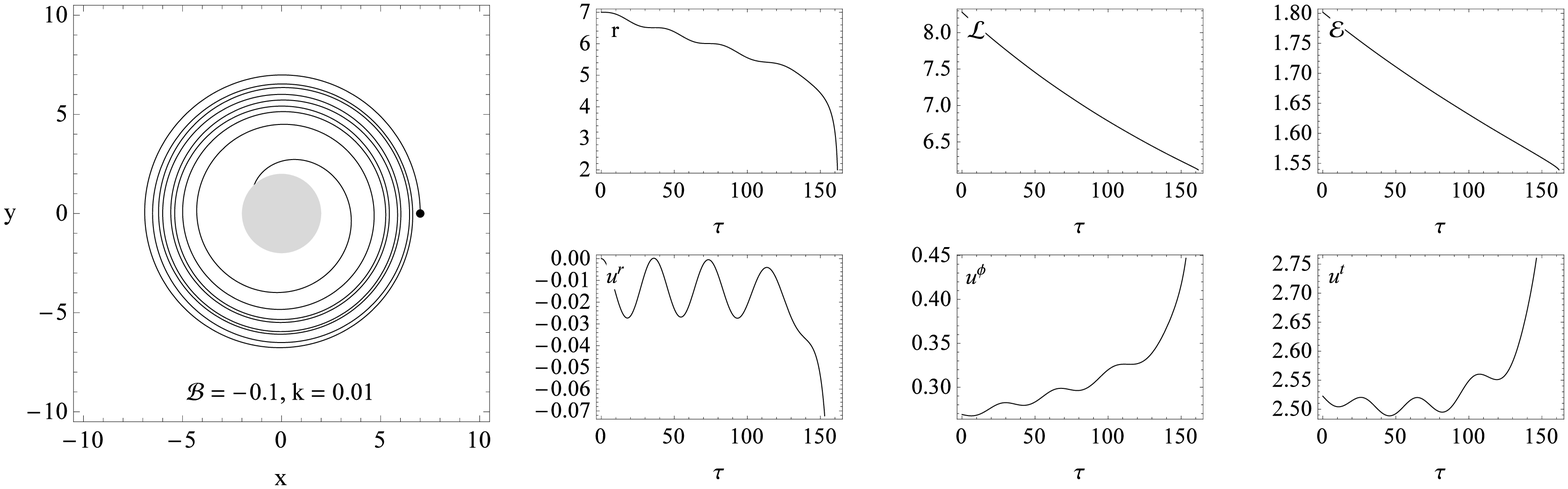}
\caption{\label{trajrad3}
Spiralling down to the black hole of charged particle due to radiation reaction of a {\it minus configuration} particle, and corresponding evolution in local time $\tau$ of the radius of orbit $r$, angular momentum $L$, energy $E$, the radial velocity $u^r$, the angular azimuthal velocity $u^\phi$ and gamma factor $u^t \equiv dt/d\tau \equiv \gamma$. The angular momentum and energy are measured with respect to observer at rest at infinity. Starting point of the particle is indicated as black dot.
}
\end{figure*}
\begin{figure*}
\includegraphics[width=\hsize]{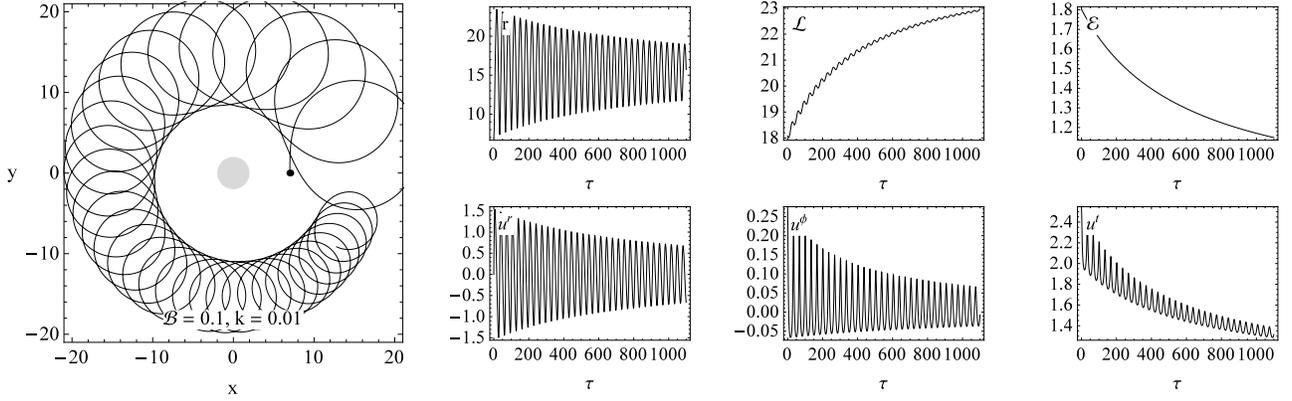}
\caption{\label{trajrad4b}
Decay of oscillations due to radiation reaction of a {\it plus configuration} particle, and corresponding evolution in time $\tau$ of the radius of orbit $r$, angular momentum $L$, energy $E$, the radial velocity $u^r$, the angular azimuthal velocity $u^\phi$ and gamma factor $u^t \equiv dt/d\tau \equiv \gamma$. The angular momentum and energy are measured with respect to observer at rest at infinity. Starting point of the particle is indicated as black dot.
}
\end{figure*}

\begin{figure*}
\includegraphics[width=\hsize]{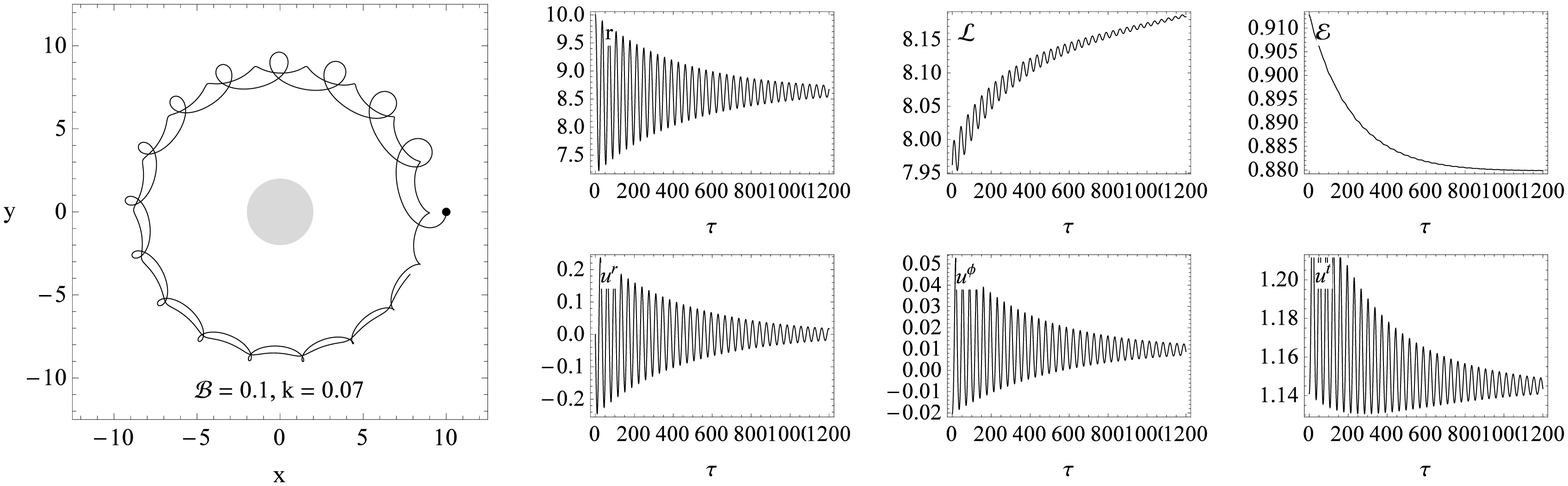}
\caption{\label{trajrad4}
Damping of oscillations due to radiation reaction of a {\it plus configuration} particle, and transition from curled motion to the nearly circular orbit. Corresponding evolution of the radius of orbit $r$, angular momentum $L$, energy $E$, the radial velocity $u^r$, the angular azimuthal velocity $u^\phi$ and gamma factor $u^t \equiv dt/d\tau \equiv \gamma$ are represented as in previous plots. The angular momentum and energy are measured with respect to observer at rest at infinity. Starting point of the particle is indicated as black dot.
}
\end{figure*}

As pointed out in Section~\ref{Section-norad}, depending on the direction of the Lorentz force one can distinguish two qualitatively different types of the motion. In a radiating case one can see such differences as well. The motion along a general worldline highly depends on the initial energy and the position of the particle. Radiation effect on the charged particle motion also has different timescales depending on whether the particle is initially oscillating or not.
The representative comparison of  trajectories of oscillating charged particle in the presence and absence of radiation reaction force is illustrated in Fig.\ref{trajrad1} for attractive Lorentz force and in Fig.\ref{trajrad2}, for repulsive one. Note that the trajectories represented in Figs. \ref{trajrad1} and \ref{trajrad2} are plotted by integration of the full form of equations of motion (\ref{cureqmo1}) and (\ref{curradforce}) without restriction to the equatorial plane.
In both cases, the motion of the charge is initially bounded. The charged particle starts its motion slightly above the equatorial plane (keeping regular character of the motion) in the vicinity of weakly magnetized Schwarzschild black hole. The initial energy and angular momentum of the particle correspond to the state close, but above the minimum of the effective potential which generates barrier where the particle bounces. Due to the action of the radiation reaction force, the charged particle changes its oscillatory character of the motion because of the loss of the energy and angular momentum. The final state of the charged particle depends on the direction of the Lorentz force. For the Lorentz force directed towards the horizon of the black hole, the synchrotron radiation leads to the collapse of the  initially stable particle into the black hole (see Fig.~\ref{trajrad1}). In the inverse case, when the Lorentz force is directed outwards the horizon (compensating thus the "gravitational attraction"), the particle remains in the bounded region in the vicinity of the black hole. On the other hand, loss of the energy of the particle leads to the disappearance of curled trajectories as well as any oscillations, as shown in Fig.~\ref{trajrad2}.

Another representative example of the dependence of the trajectories on the alignment of the external magnetic field is shown in Figs.~\ref{trajrad3} and \ref{trajrad4b}, where the initial conditions differ only in the sign of the magnetic parameter $\BB$, while the other parameters are chosen to be the same. As before, for the attractive Lorentz force, the charged particle spirals down to the black hole (Fig.\ref{trajrad3}), while in the repulsive case, the motion is stable (Fig.\ref{trajrad4b}).
Stability of the {\it plus configurations} against collapse can be explained due to the reason that a radiating particle is assumed to be accelerated by the external Lorentz force only. This implies that the radiation reaction in fact decreases the gyroradius of the radiating particle. In the case when the Lorentz force is directed outwards the black hole, the gyrocenter of oscillating charged particle is located near its orbit (center of "curls"), while in the case of {\it minus configuration}, the gyrocenter of the orbit is located inside the horizon (it coincides with a black hole singularity for the circular motion). Indeed, from the analogy with the flat spacetime formula (\ref{ladeq}), the sign of the radiation reaction force depends on the alignment of the Lorentz force due to the reduction of order procedure performed in the beginning. Thus, the "Lorentz-repulsive" case with radiation reaction represents damping of oscillations, turning the oscillatory epicyclic type of motion to nearly circular orbit. A strong non-linearity of equations of motion shifts the location of stable circular orbits as well, however, as we will see below, this effect is relatively slow in comparison to the process of radiative damping of oscillations. One can conclude that for relatively short periods of time an oscillating charged particle, undergoing repulsive Lorentz force and radiation reaction force will damp its oscillations settling down to the circular orbit. This result is represented in Fig.\ref{trajrad4}, being in accord with the qualitative predictions given in \cite{Shoom:2015:PHYSR4:}. 

\subsection{Evolution of energy and angular momentum} \label{subsection-E-L-tau}

In fact, the synchrotron radiation carries out the kinetic energy of the particle. One can see in $E-\tau$ plots of Fig.\ref{trajrad3} and Fig.\ref{trajrad4b}, that the energy slightly decreases in both cases. However, the azimuthal angular momentum of the particle changes differently in dependence on the direction of the Lorentz force. 
While in the {\it minus configurations} the charged particle decreases its angular momentum due to radiation, in the {\it plus configurations} the angular momentum quasi-periodically increases. Quasi-periodicity is connected with the quasi-harmonic oscillations of the particle in the motion with "curls". Increase of the angular momentum of the particles in the {\it plus configurations} becomes clear if one compares the angular momentum of charged particles in the motion with curls, $\cl_{*}$ given by (\ref{curl-L}), with those of the circular orbits given by \cite{Kol-Stu-Tur:2015:CQG:}
\beq \label{L-cir-orbit}
\cl_{\rm c} = \frac{-\BB r^2 + r \sqrt{\BB^2 r^2 (r-2)^2 + r - 3}}{r-3}.
\eeq
Let us now assume that the initial angular momentum of the particle is $\approx \cl_{*}$, while the angular momentum at the final state is nearly equal to those of the stable circular orbit, $ \approx \cl_{\rm c}$. From the fact that that $\cl_{\rm c} > \cl_{*}$ (see details in \cite{Kol-Stu-Tur:2015:CQG:}) one can conclude that the particle is actually gaining an angular momentum by reducing radial oscillations due to radiation reaction force.
In the {\it minus configurations}, the trajectories with curls, i.e. the regions with negative angular velocity cannot be observed. Therefore, a charged particle initially located at a bounded orbit around a black hole continuously loses its angular momentum due to radiation reaction force. When angular momentum of the particle becomes lower than the one corresponding to the innermost stable circular orbit (ISCO), $\cl < \cl_{\rm ISCO}$, the particle collapses to the black hole. The value of angular momentum at ISCO is given by \cite{Kol-Stu-Tur:2015:CQG:}
\beq
\cl_{\rm ISCO} = - 2 \BB r_I + \sqrt{\BB^2 r_I^2 (5 r_I^2 - 4 r_I + 4) + 2 r_I},
\eeq
where $r_I$ is the ISCO radius. Detailed analysis of the boundaries of angular momentum separating different types of orbits in the vicinity of magnetized black holes can be found in \cite{Kol-Stu-Tur:2015:CQG:} for non-rotating black holes and in \cite{Tur-Stu-Kol:2016:PHYSR4:} for magnetized Kerr black hole.

\subsection{Lifetime of oscillations} \label{t-damp-r}

\begin{figure}
\includegraphics[width=0.8\hsize]{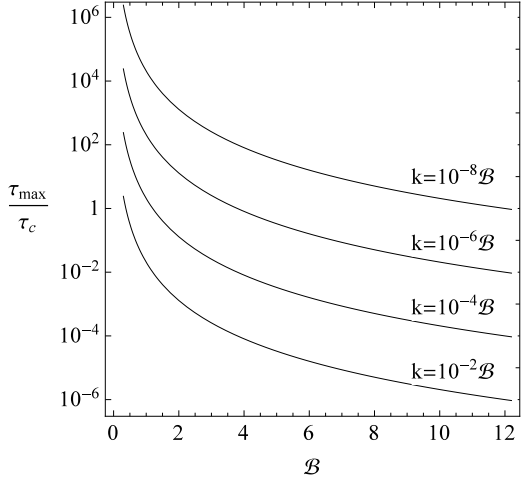}
\caption{\label{taumax-tauco}
Number of revolutions of the oscillating charged particle around black hole given as a ratio of the maximal decay time to the orbital time in dependence on the magnetic parameter $\BB$ and different values of radiation parameter $k$.
}
\end{figure}

\begin{figure*}
\includegraphics[width=\hsize]{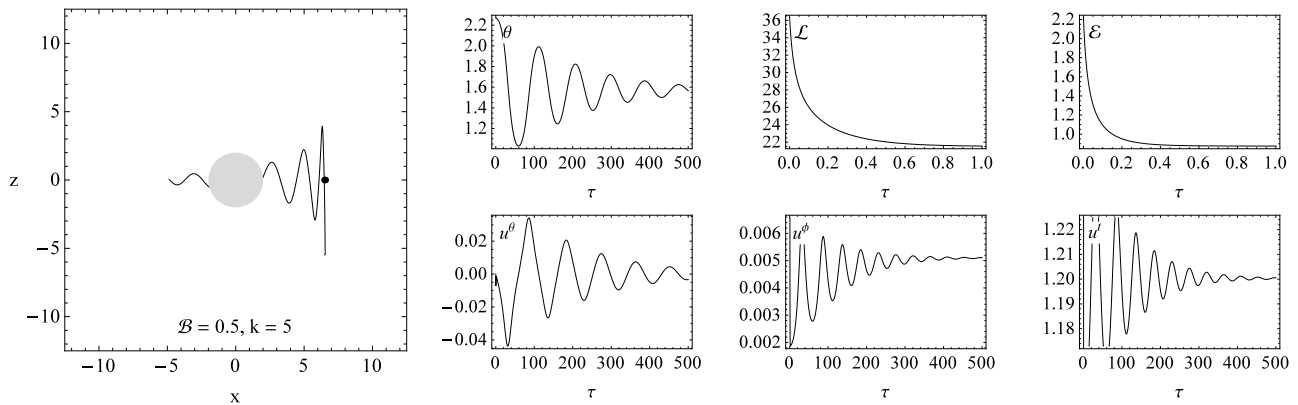}
\caption{\label{trajrad6}
Damping of vertical oscillations of a charged particle revolving near the circular orbit of a black hole in magnetic field and corresponding evolution in time $\tau$ of the angle $\theta$, angular momentum $L$, energy $E$, the vertical velocity $u^\theta$, the angular azimuthal velocity $u^\phi$ and gamma factor $u^t \equiv dt/d\tau \equiv \gamma$. The angular momentum and energy are measured with respect to observer at rest at infinity. Starting point of a particle is indicated as black dot.
}
\end{figure*}

One can calculate the relaxation time ${\tau}$ of a charged particle required for decay of the radial oscillations due to radiation reaction force. Calculation of the relaxation time makes sense in the {\it plus configurations} only, when the radiation reaction does not cause the particle collapse to the black hole. The rate of the energy loss can be written as 
\beq
\dot{\ce} = \frac{\ce_f - \ce_i}{\tau},
\eeq
where $\ce_i$ and $\ce_f$ are the initial and final energies of the particle. For the particle with velocity close to the speed of light, $v \sim c$, the leading contribution for the energy loss is given by the first term of the expression (\ref{enerlossSCH}), which can be solved analytically giving
\beq \label{ultraenergyloss}
\frac{d \ce}{d\tau} = - 4 \BB^2 k \ce^3, \quad \ce(\tau) = \frac{\ce_i}{\sqrt{1 + 8 \BB^2 k \ce^2_i \tau}}.
\eeq
Extracting $\tau$ from (\ref{ultraenergyloss}) at the final energy, we get
\beq \label{relaxtau1}
\tau = \frac{1}{4 k \BB^2} \frac{\ce^{2}_{i} - \ce^{2}_{f}}{ \ce^{2}_{i} \ce^{2}_{f} }.
\eeq
One can find the upper limit of the relaxation time required to lower the maximal energy at the bounded orbit to the lowest energy corresponding to the circular orbit, where no oscillations exist. Thus, identifying $\ce_i = \ce_+$ and $\ce_i = \ce_-$, given by (\ref{trappedenergy}), and assuming locations of extrema $ r^E_{\rm max} \approx r^E_{\rm min} = r$ (indeed, according to \cite{Kol-Stu-Tur:2015:CQG:}, for large values of $\BB$ the locations of extrema are very close), we get
\beq \label{max-lifetime1}
\tau_{\rm max} = \frac{\sqrt{r-3+\BB^2 f^2 r^4}}{k \BB \left(1 + 4 \BB^2 r^2 \right) f^2}.
\eeq
Note that this equation corresponds to the particles with ultrarelativistic velocities. This implies that for large values of magnetic parameter $\BB$, one can write Eq.(\ref{max-lifetime1}) in the following simple form
\beq \label{max-lifetime2}
\tau_{\rm max} \approx \frac{1}{k \BB^2 f(r)}, \quad \BB \gg 1,
\eeq
where $f(r) = 1-2 M/r$ is the lapse function. In particular, this equation shows that closer to black hole, the decay of oscillations of charged particle is slower.

It is useful to compare the maximal relaxation time $\tau_{\rm max}$ of the charged particle oscillations with the orbital time of a charged particle around black hole. The orbital frequency $\omega_{\phi}$ is defined in Eq.(\ref{freq-omK-omB}). Considering the stable circular orbits at the equatorial plane, we get the time of one revolution around the black hole in the form
\beq \label{time-cir-or}
\tau_{\rm c} =  \frac{2\pi r^2}{\cl_{\rm c} - \BB r^2},
\eeq
where $\cl_{\rm c}$ is given by (\ref{L-cir-orbit}). Dividing (\ref{max-lifetime1}) by (\ref{time-cir-or}), we get the maximal number of revolutions around black hole during which the radial oscillations decay
\beq 
 N_{\rm max}  \equiv  \frac{\tau_{\rm max}}{\tau_{\rm c}}  =  \frac{ \BB r A }{\pi k (r-2)^2 \left(4 \BB^2 r^2 + 1 \right) (\BB r (r-2)+A)},
\eeq
where $A = \sqrt{r-3+\BB^2 r^2(r-2)^2}$. The dependence of $N_{\rm max}$ on the magnetic parameter $\BB$ is given in Fig.\ref{taumax-tauco}, for different values of the radiation parameter $k$. Since both the parameters $\BB$ and $k$ depend on the specific charge of the charged particle, the parameters are not really independent. This implies that the parameter $k$ can be expressed as fraction of the parameter $\BB$. As one can see from Fig.\ref{taumax-tauco}, the radiation reaction of the charged particle cannot be neglected even for small values of the parameter $k$. For large values of $\BB$, initially oscillating charged particle may decay its radial oscillations ending up at the circular orbit pathing only fraction of the total orbit around the black hole.
The estimations of both the parameters $\BB$ and $k$, and the lifetime of the charged particle oscillations in realistic scenarios for electrons and ions will be given in Sec.~\ref{astro-section}.

Equations of motion (\ref{cureqmo1}) and (\ref{curradforce}) allow us to repeat the calculations of the damping of vertical oscillations (in $\theta$) of charged particles as well. In the case when the initial velocity of a particle is ultrarelativistic, the leading term responsible for the radiation reaction for all components of velocities is the last term of Eq.(\ref{curradforce}). This implies that the decay rate of the vertical oscillations of charged particle is similar to the decay rate of the radial oscillations. The representative plots of relevant parameters of charged particles during decay of vertical oscillations is presented in Fig.\ref{trajrad6}.

\subsection{Circular orbit widening} \label{section-shift}

\begin{figure}
\includegraphics[width=0.8\hsize]{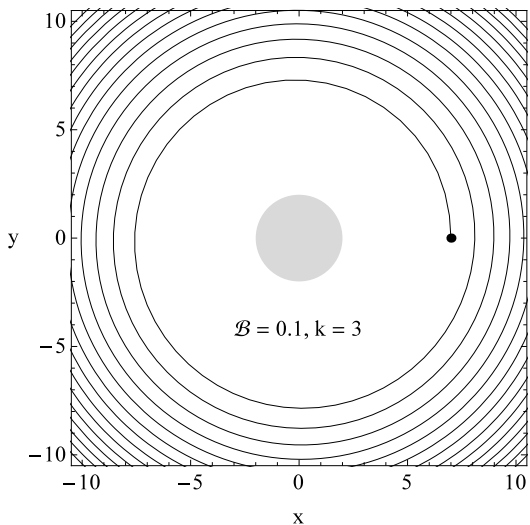}
\caption{\label{trajrad5}
Radiative widening of the stable circular orbit of a charged particle due to radiation reaction force. Starting point of a particle is indicated as black dot.
}
\end{figure}
\begin{figure}
\includegraphics[width=0.8\hsize]{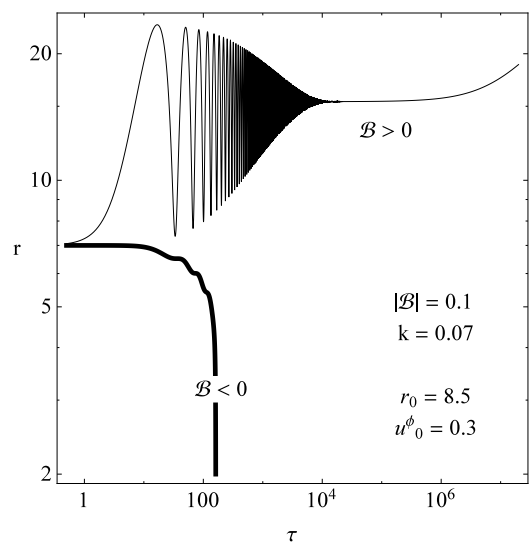}
\caption{\label{trajrad7}
Evolution of radial position of charged particle in time for two configurations corresponding to trajectories illustrated in Figs \ref{trajrad3} and \ref{trajrad4b}.
}
\end{figure}

Another new consequence of the radiation reaction force in the vicinity of a black hole and in presence of external magnetic field is the evolution of the circular orbits of charged particles during the radiation process and shifting of the radius of circular orbit outwards from the black hole. Obviously, the last case can be realized for repulsive Lorentz force only, while in attractive case, as we have already seen, the particle spirals down to the black hole and no closed circular orbits can be formed. Let us consider a purely circular motion of a charged particle revolving around black hole at the equatorial plane, in presence of uniform  magnetic field (see Fig.\ref{trajrad5}). While small oscillations can appear, they will be dumped relatively fast and the energy of the particle can be considered to be always located near the minimum of the effective potential corresponding to the circular orbits. The radiation reaction force in the frame moving with the particle vanishes, however, in the locally geodesic frame of reference, the self force is directed anti-parallel to the particle velocity and concentrated in the narrow cone along the motion. This implies that, in fact, the radiation reaction reduces the angular velocity of the particle. Reducing the angular velocity while keeping the circular character of the motion causes the increasing of the radius of the circular orbit. This, in turn, increases the particle energy and angular momentum with respect to the observer at rest at infinity. Indeed, the energy and angular momentum of the charged particle for an observer at infinity are given as
\bea
\ce = \left(1-\frac{2}{r}\right) u^t, \\
\cl = r^2 \left(u^{\phi} + \BB\right),
\eea
which are not conserved for radiating particles. When the radius of the orbit $r$ increases faster than the deceleration of velocities $u^t$ and $u^{\phi}$, then the energy and angular momentum of the charged particle will increase accordingly. Example of the widening of the circular orbit of radiating charged particle is illustrated in Fig.\ref{trajrad5}. Note that the parameter responsible for the radiation reaction $k$ is taken to be unrealistically large in order to obtain representative plot, since for small values of the parameter $k$ due to slowness of the process the shift of the orbit would be barely seen. 

The process of the expansion of circular orbits is much slower in comparison to the process of decay of the oscillations. Representative changes of the positions of charged particles in time for attractive and repulsive Lorentz forces are demonstrated in Fig.\ref{trajrad7}. In the case of the attractive Lorentz force ($\BB<0$), the charged particle plunges to the black hole, as described in the previous section. The wiggles in ${\cal B} < 0$ case appear due to the precession of the particle around circular orbit, as can be also seen in $r-\tau$ plot of Fig.\ref{trajrad3}. In the repulsive case, the particle starts to oscillate inside a barrier given by the effective potential at a given moment of time. Due to the radiation reaction force, it decays its oscillations ending up at a circular orbit with larger radius than the initial position. At the circular orbit, particle continues to slow down its orbital velocity, keeping the stable motion which affects on the location of the circular orbit, shifting it towards infinity. 
Note that the timescale of the plot is logarithmic, which implies that the process of shifting of the particle circular orbits is relatively slow.

The energy loss of this process is given by (\ref{enerlossSCH}), where all terms are important since the motion is not necessarily ultrarelativistic.
Since all quantities in (\ref{enerlossSCH}) are implicit functions of time, one can rewrite this equation in the form
\beq \label{enerlossSCH-full}
{\dot{\ce} } = - A_{1} \ce^3 + A_{2} \ce x(\tau),
\eeq
where dot denotes the derivative with respect to the proper time $\tau$, $A_1 = 4 k \BB^2$ and $A_2 = 2 k \BB$ are constants and  $x(\tau) = 2 k \BB f (\tau) + u^{\phi} (\tau) / r(\tau)$. The analytical form of the solution is found to be
\beq \label{wideningenergy1}
\ce (\tau) = \frac{\ce_i e^{A_2 X(\tau)}}{\left(1 + 2 A_1 \ce_i^2 \int_0^{\tau} e^{2 A_2 X(\tau')} d\tau'\right)^{\frac12} },
\eeq
where $X(\tau) = \int_0^\tau x(\tau) d\tau$.
The equation (\ref{enerlossSCH-full}) can be solved numerically for the given set of equations of motion, and it is performed for all $\ce-\tau$ dependence plots of the present paper. The detailed analysis of the effect of radiative widening of orbits is left for our future studies.

\section{Relevance to astrophysics} \label{astro-section}


\begin{table}[]
\begin{center}
\begin{tabular}{c || c | c   }
\hline
B (Gauss)    &  $\tau_e$ (s)  &  $\tau_p$ (s)     	  \\	 										
\hline \hline 

$10^{12}$	   &  ~~~$10^{-16}$~~~ & ~~$10^{-6}$~~   \\	
$10^{8}$	   &  ~~~$10^{-8}$~~~ & ~~$10^{2}$~~   \\				
$10^{4}$	   &  ~~~$1$~~~       & ~~~$10^{10}$~~~   \\				
$1$	         &  ~~~$10^{8}$~~~ & ~~~$10^{1}$~~~   \\
$10^{-4}$   &  ~~$10^{16}$~~   & ~~$10^{26}$~~     \\						

\hline
\end{tabular}
\caption{Typical decay times of electrons $\tau_e$ and protons $\tau_p$ orbiting a black hole for different values of magnetic field $B$.
\label{tab1}
}
\end{center}
\end{table}

In order to relate our results to realistic astrophysical scenarios, we perform the estimations of relevant parameters of the discussed model. 
Even thought the magnetic field does not violate the geometry of the background spacetime, satisfying the condition (\ref{BBB}), one cannot neglect its effect on the motion of charged particles due to the large values of the specific charge (charge per mass ratio) for elementary particles. Restoring the world constants, the dimensionless parameter $\BB$, widely used in the present paper, takes the form
\beq \label{BB-param}
\BB = \frac{|q| B G M}{2 m_e c^4},
\eeq
reflecting the relative influence of the gravitational and magnetic fields on the charged particle motion. According to \cite{Pio-etal:2011:ASBULL:,Baczko-etal:2016:AAP:}, the characteristic values of the magnetic fields near the stellar mass and supermassive black holes are $B \sim 10^8 {\rm G}$, for $M = 10 M_{\odot}$ and $B \sim 10^4 {\rm G}$, for $M = 10^9 M_{\odot}$. For electrons, one can estimate the parameter $\BB$ as
\bea 
\BB_{\rm BH} \approx 4.32 \times 10^{10} \quad {\rm for} \,\,\, M = 10 M_{\odot}, \label{estimation-BB1}   \\
\BB_{\rm SMBH } \approx 4.32 \times 10^{14} \quad {\rm for} \,\,\, M = 10^9 M_{\odot}. \label{estimation-BB2} 
\eea
As a representative example one can estimate the parameter $\BB$ for an electron in the vicinity of the supermassive black hole (SMBH) at the center of the Milky Way, which is currently the best studied SMBH candidate. The equipartition strength of the magnetic field near the Galactic Center is usually considered to be of tens of Gauss (\cite{Johnson-etal:2015:Science:}). Moreover, multi-frequency measurements of pulsar near the Galactic Center by \cite{Eatough-etal:2013:Natur:} demonstrate the existence of strong magnetic field of few hundred Gauss near the event horizon of SMBH. The mass of the black hole candidate is measured to be $4.3 \times 10^6 M_{\odot}$ (for more details about Sgr A*, see recent review by \cite{Eckart-etal:2017:FOP:}). Thus, the parameter $\BB$ for the electron surrounding SMBH at the center of our Galaxy can be estimated as
\beq \label{estimation-BBsgra}
\BB_{\rm SgrA^*} \approx \frac{|e| B G M}{2 m_e c^4} \approx 1.86 \times 10^{10},
\eeq
For protons, the values of $\BB$ in (\ref{estimation-BB1}) - (\ref{estimation-BBsgra}) are lower by the factor $m_p/m_e \approx 1836$.
Large values of the parameter $\BB$ in realistic scenarios imply that the effects of magnetic field on the dynamics of charged particles play one of the essential roles.

Radiation reaction force acting on a charged particle is represented by the dimensionless parameter $k$ which has the form
\beq \label{k-param}
k = \frac{2}{3} \frac{q^2}{m G M}.
\eeq
The value of parameter $k$ is much lower than those of $\BB$. For electrons orbiting stellar mass and supermassive black holes we have respectively
\bea \label{estimation-k1}
k_{\rm BH} \sim 10^{-19} \quad {\rm for} \,\,\, M = 10 M_{\odot},  \\
k_{\rm SMBH } \sim 10^{-27} \quad {\rm for} \,\,\, M = 10^9 M_{\odot} .
\eea
For electron around Sgr A*, we have $k_{\rm SgrA^*} \sim 10^{-25}$. For protons, the values of $k$ parameter is lower by the factor $m_p/m_e \approx 1836$ as in case with $\BB$. Despite the weakness of parameter $k$ as compared to $\BB$, it enters into the equations for ultrarelativistic particles as $k\BB^2$, which can make the effect of the radiation reaction force considerably large.

One can estimate the timescale of the decay of charged particle oscillations. Typical orders of magnitude of the oscillation decay time of an electron and proton orbiting black hole are given in Tab.\ref{tab1}. One can compare these values with the timescale of one orbit of particle $\tau_c$ around stellar mass and supermassive black holes. For the orbit at ISCO we have
\bea
\tau_c \sim 10^{-3} {\rm s}, \quad {\rm for} \,\,\, M = 10 M_{\odot},  \\
\tau_c \sim 10^{4} {\rm s}, \quad {\rm for} \,\,\, M = 10^9 M_{\odot},
\eea
For Sgr A* the electron decay time is about $10^4$~s, while orbiting time at ISCO is $\sim 10^3$~s. It is interesting to note that for ions the decay time of oscillations is much less than for electrons, namely for the factor of $(m_p/m_e)^3 \sim 10^{10}$. Thus, one can conclude, that the radiation reaction of electrons can be quite relevant for the astrophysically plausible magnetic fields providing reasonable decay times, while radiation reaction of ions can be relevant only in the presence of magnetic fields much stronger than those corresponding to black holes, e.g., in the vicinity of neutron stars.

Estimations of the timescale of the radiative widening of charged particle orbit
 show that this process is slower than the damping of oscillations due to radiation reaction for about $10^{10}$ times in average, for the values of parameters $\BB$ and $k$ given by equations (\ref{estimation-k1}) and (\ref{estimation-BB1}). However in some astrophysical scenarios with large magnetic fields this process can be potentially measured.

The radiation reaction and related decay time of particle oscillations can have a significant effect in plasma surrounding black hole, when the decay time of charged particle becomes smaller than the average time between collisions of particles in plasma.

\section{Summary} \label{conclusion}

We have studied the radiation reaction of charged particles moving around Schwarzschild black hole immersed into external asymptotically uniform magnetic field. 
We started analysis from the study of the motion in flat spacetime, described by the Lorentz-Dirac equation and by its reduced form -- the Landau-Lifshitz equation. 
We have tested both equations numerically and found that in the simplified framework of the asymptotically uniform magnetic field both approaches lead to the same result. The Landau-Lifshitz equation is more convenient as it is a second order differential equation which can be solved in usual manner, while for the Lorentz-Dirac equation we have to perform the integration of dynamical equations backwards in time, in order to avoid the exponential increase of computational error. 

We have shown that the tail term appearing in the general formalism for the motion in curved spacetime can be neglected in the presence of magnetic field. Vanishing the Ricci term for Schwarzschild black hole case enables to use the covariant form of the Lorentz-Dirac equation or its reduced form -- the covariant Landau-Lifshitz equation.
The motion of charged particles around non-rotating black hole in the presence of external uniform magnetic field can be classified into two different types, differing by the orientation of the Lorentz force directed towards and outwards from the black hole, which we called as {\it minus} and {\it plus configurations}, respectively. We concentrated attention on the quasi-bounded orbits characterized by the presence of maxima and minima of the effective potential. In such orbits, the charged particles can undergo the stable quasi-harmonic oscillations in the radial and vertical directions. We have shown that the presence of the radiation reaction force leads to the decay of the particle oscillations, while the decay time increases as the orbit approaches to the black hole. The final state of the particle depends on the orientation of the Lorentz force with respect to the black hole. In case when the Lorentz force is directed towards the black hole, the radiation reaction leads to the fall of the charged particle from initially stable orbit into the black hole. Inversely, when the Lorentz force is repulsive, the orbit of the charged particle remains bounded while oscillations decay. 

We have shown that in the absence of oscillations, the radiation reaction force can shift the circular orbits outwards from the black hole. This can occur only in case when the Lorentz force is repulsive. While the stability of the circular orbit is conserved, the radiation reaction acting against the particle motion leads to decreasing of the linear velocity of the particle which in turn leads to the widening of the orbit. Even though the kinetic energy of a particle decreases due to the synchrotron radiation, its potential energy increases faster due to the widening of the orbit. One can find efficiency of this process defined as the ratio between the gain energy required for the shifting the particle from ISCO to infinity to its final energy. The value of the efficiency depends on the initial position of the particle and the parameters $\BB$ and $k$, which are governing the magnetic field and radiation reaction force, respectively. The efficiency can reach the values, close to the maximal $100\%$. 
The maximal efficiency, however can never be achieved, since it would require $\BB \rightarrow \infty$, and the ISCO would coincide with the black hole horizon. In the absence of magnetic field, the formal efficiency of mechanical process of shifting the particle orbit from infinity to the ISCO is $5.7\%$. One needs to note that the process of widening of orbits is many orders of magnitude slower that the process of cooling of the particle oscillations. The physical mechanism causing the gain in energy requires further investigations.

We have estimated the relevant parameters of the model as well as the decay times of the particle oscillations for typical values of the magnetic field for stellar mass and supermassive black holes. Assuming the test particle as an electron, we have found that the values of the decay times are reasonable and can be measured in most of the astrophysical scenarios. Despite to the slowness of the process of the radiative widening of orbits the estimations show that for stellar mass black holes in the presence of magnetic fields of order $10^8$G and more, the widening of orbits can be potentially measured.

We expect that our results can be relevant in treating variety of astrophysical phenomena observed in microquasars or active galactic nuclei, containing stellar mass or supermassive black holes. These results can put relevant limits on validity of recent models of high-frequency quasi-periodic oscillations (HF QPOs), or motion of jets observed in microquasars. The limits implied by the radiation reaction forces can be clearly relevant for both the magnetized geodesic models of HF QPOs discussed in \cite{Kol-Tur-Stu:2017:arXiv:}; see also \cite{Stu-Kot-Tor:2013:AAP:}, or the string loop models of jets or HF QPOs discussed in \cite{Jac-Sot:2009:PHYSR4:,Stu-Kol:2012:JCAP:,Kol-Stu:2013:PHYSR4:,Tur-etal:2014:PRD:}.

\section*{Acknowledgments}

%
A.T. and M.K. acknowledge the Czech Science Foundation Grant No. 16-03564Y and the Silesian University in Opava Grant No. SGS/14/2016.
Z.S. acknowledges the Albert Einstein Centre for Gravitation and Astrophysics supported by the Czech Science Foundation Grant No. 14-37086G.
D.G. acknowledges the support of the Russian
Foundation of Fundamental Research Grant 17-02-01299a
and the Russian Government Program of Competitive Growth of the
Kazan Federal University.





\end{document}